%Paper: hep-ph/9401249
%From: U6404939@ucsvc.ucs.unimelb.edu.au
%Date: Fri, 14 Jan 1994 10:43:46 +1000

\magnification=1200
{\nopagenumbers
\baselineskip=15pt
\hsize=5.5in
\hskip10cm UM--P-94/01\hfil\par
\hskip10cm OZ-94/01 \hfil
\vskip2cm
\centerline{\bf Calculating Fragmentation Functions from Definitions}  \par
\vskip2cm
\hskip4cm J.P. Ma \par
\vskip0.5cm
\hskip4cm Recearch Center for High Energy Physics \par
\hskip4cm School of Physics\par
\hskip4cm University of Melbourne \par
\hskip4cm Parkville, Victoria 3052\par
\hskip4cm Australia \par
\vskip2cm
{\bf\underbar{Abstract}}:\par
Fragmentation functions for hadrons composed of heavy quarks are calculated
directly from the definitions given by Collins and Soper and are compared
with those calculated in another way. A new fragmentaion function for
a P-wave meson is also obtained and the singularity arising at the
leading order is discussed.
\par\vfil\eject}

%begin the text
\baselineskip=20pt
\pageno=1
Recently it has been shown[1,2,3,4]  that
 the fragmentation functions for a hadron
as a bound-state of two heavy quarks can be calculated perturbatively. The
reason such calculations are possible is that
the bound-states can be described well
 by the nonrelativstic wavefunctions and hence the effect of long distance
in the fragmentaion can be factorized into the wavefunction, expressed
through the radial wavefunction at the origin. Although it is realized[2,4]
that the fragmentaion functions are universal, the fragmentation functions
calculated in these works are extracted from specific processes. It should
be noted that in QCD the definitions of the fragmentaion functions (or
decay functions) are already given by Collins and Soper[5] and these
functions should be universal and independent of pocesses. The so called
factorization thorem is based on such definitions(for the factorization
theorem see [6] and references cited there). It is interesting to
calculate fragmentation functions directly from these defintions
and to compare those obtained previously. On the other hand, it is also
convenient to work with the defintions for calculating the high order
corrections.
In this letter we will calculate fragmentation functions
from the definitions.
\par
To give the definitions for a fragmentaion function it is convenient
to work in the light-cone coordinate system. In this coordinate system
a 4-vector $p$ is expressed as $p^{\mu}=(p^+,p^-,{\bf p_T})$, with
$p^+=(p^0+p^3)/\sqrt{2},\ p^-=(p^0-p^3)/\sqrt{2}$.
Introducing a vector  $n$ with
$n^{\mu}=(0,1,{\bf 0_T})$, the fragmentation functions
for a spinless hadron $H$ are defined
as[5]:
  $$\eqalign { D_{H/Q}(z)=& {z\over 4\pi} \int dx^- e^{ -iP^+x^-/z}
    {1\over 3}{\rm Tr_{color}}{1\over 2}{\rm Tr_{Dirac}}\{
    n\cdot \gamma <0\vert Q(0) \cr
     & \ \ \ \  \bar P {\rm exp}\{-ig_s
    \int_0^\infty d\lambda n\cdot G^T(\lambda n^{\mu}) \}
    a_H^\dagger (P^+, {\bf 0_T}) a_H(P^+,  {\bf 0_T}) \cr
  & \ \ \   P {\rm exp}\{ ig_s \int_{x^-}^\infty d\lambda
        n\cdot G^T(\lambda n^{\mu})
   \} \bar Q (0, x^-,{\bf 0_T}) \vert 0 >  \cr
     D_{H/G}(z) =& {-z \over 32 \pi k^+} \int dx^- e^{ -iP^+x^-/z}
     <0 \vert G^{b, +\nu}(0) \cr
     & \ \ \ \  \{\bar P {\rm exp}\{-ig_s
    \int_0^\infty d\lambda n\cdot G(\lambda n^{\mu}) \} \}^{bc}
    a_H^\dagger (P^+, {\bf 0_T}) a_H(P^+,  {\bf 0_T}) \cr
   & \ \ \  \{ P {\rm exp}\{ ig_s \int_{x^-}^\infty d\lambda
      n\cdot G(\lambda n^{\mu})
   \} \}^{cd} G^{d,+}_{\ \ \nu}(0,x^-, {\bf 0_T}) \vert 0> \cr }
    \eqno(1) $$
Where $G_{\mu}(x)=G^a_{\mu}(x) {\lambda ^a \over 2}$, $G_{\mu}^a (x)$ is the
gluon field and the $\lambda ^a (a=1,\dots, 8)$ are the Gell-Mann matrices.
The subscription $T$ denotes the transpose.
$G^{a,\mu\nu}$ is the gluon field strength and $a^\dagger _H({\bf P})$
is the creation operator for the hadron $H$. The function
 $ D_{H/Q}(z)$ or $D_{H/G}(z)$ are interpreted as the probablity
 of a quark $Q$ or a gluon $G$ with momentum $k$ to decay into the hadron $H$
with momentum component $P^+=z k^+$, both are gauge invariant from the
definitions. The defintions in (1) are
the unrenormalized versions, i.e. all quantities are bar quantities(for
renormalization see [5]). When one has a parton, i.e. quark or gluon,
instead of the hadron in (1),
the perturbative expansion in $g_s$, the strong coupling constant,
can also be expressed with Feymann
diagrams, the Feymann rules can be found in [5,6]. The definitions can easily
be generalized to hadrons with nonzero spin.
\par
{}From the defintions a direct calculation is possible if one can express
the hadron operators with the parton operators. This is possible for
a hadron composed of two heavy quarks, where, as mentioned, the
nonrelativstic approach works. To see how to express the hadron operator
as a parton one, let us consider a spinless hadron to be a bound-state
of a heavy quark $Q_1$ and an anti-quark $\bar Q_2$ and to have the quantum
number $^1S_0$. In
its rest frame the state can be expressed as:
  $$ \vert  ^1 S_0 > = A_0 \int {d^3 q \over (2 \pi)^3}
      f_0 (\vert {\bf q}\vert ) Y_{00}(\theta ,\phi)
         \chi_{s_1 s_2} {1\over \sqrt{3}}\sum_{color}
       a^\dagger _{s_1} ({\bf q})
      b^\dagger _{s_2}(-{\bf q}) \vert 0> \eqno (2) $$
with $$ \chi_{s_1s_2}={1\over \sqrt{2}} \left( \matrix{ 0 &1 \cr
                                -1& 0 \cr}\right)_{s_1s_2}\eqno(3)$$
In Eq(2),  $a^\dagger _{s_1}(b^\dagger_{s_2})$
 is the creation operator for $Q_1(\bar Q_2)$,
the indices $s_1$ and $s_2$ are spin-indices, and $A_0$ is a suitable
normalization constant.
The function $\psi_0 ({\bf q})= f_0(\vert {\bf q}\vert) Y_{00}(\theta ,\phi)$
is obtained
through Fourier transformation of the nonrelativstic wavefunction.
$f_0(\vert {\bf q}\vert)$ goes rapidly to zero, when $\vert {\bf q}\vert
\rightarrow \infty$. This means, the region with $\vert {\bf q}\vert\approx
 0$ is dominant in the integral of Eq.(2). In this work we will always
take nonzero leading order in $\vert {\bf q}\vert$ and the hadron mass
 $M=m_1+m_2$.
Through a Lorenz boost
the state can be transformed into a ${\bf P} \not= 0$ state:
  $$ \vert ^1 S_0, {\bf P} > = A_0 \int {d^3 q \over (2 \pi)^3}
      f_0 (\vert {\bf q}\vert ) {1\over \sqrt {4\pi}} \chi_{s_1 s_2}
          {1\over \sqrt {3}}\sum_{color} a^\dagger _{s_1} ({\bf p_1})
      b^\dagger _{s_2}({\bf p_2}) \vert 0> \eqno (4) $$
where
   $$ p_1= {m_1 \over m_1+m_2} P +q_p,\ \ p_2={m_2 \over m_1+m_2}P
 -q_p \eqno (5) $$
$q_p$ is related through the Lorenz boost to $q$ in the rest frame.
Now it is easy to show that the bound-state satisfies the normalization
condition given in [5] as required for the definitions in (1), if one takes
the following form for the hadron creation operator:
  $$ \eqalign { a^\dagger _H(P) =&  A_0
   \int {d^3 q\over (2\pi)^3}
    f_0 (\vert {\bf q}\vert ) {1\over \sqrt {4\pi}} \chi_{s_1 s_2}
         {1\over \sqrt{3}}\sum_{color}  a^\dagger _{s_1} ({\bf p_1})
      b^\dagger _{s_2}({\bf p_2})\cr
        A_0^{-1}=& \sqrt{{2m_1m_2\over m_1+m_2}\cdot {P^0+P^3 \over P^0+
     \vert {\bf P}\vert } }\cr}     \eqno(6) $$
Similarly, one can also obtain the operator for the $^3S_1$ hadronic state:
   $$  a^\dagger _H(P,\lambda ) =A_0
    \int {d^3 q\over (2\pi)^3}
    f_0 (\vert {\bf q}\vert )  {1\over \sqrt {4\pi}}
      ( \epsilon_i (\lambda)\cdot \sigma _i
      \chi)_{s_1 s_2}
       {1\over \sqrt{3}}\sum_{color} a^\dagger _{s_1} ({\bf p_1})
      b^\dagger _{s_2}({\bf p_2})     \eqno(7) $$
Here $\sigma_i (i=1,2,3)$ are the Pauli matrices and $\epsilon_i(\lambda)
 (i=1,2,3)$ is the component of the polarization vector of the hadron
 in its rest frame. $\lambda$ labels the heilicity with respect to the
 ${\bf P}$ direction.
With Eq.(6) and Eq.(7) we can now calculate the fragmentaion function
for hadrons with quantum number $^1S_0$ and $^3S_1$ from the definions
in Eq.(1). In the leading order of $g_s$ there are four Feymann diagrams
for a quark $Q_1$ to decay to those hadrons. The diagrams are given in Fig.1,
 where
the double lines represent the Wilson line operators in Eq.(1).
One can work in the light cone gauge $n\cdot G=0$. In this gauge the
Wilson line operators in the defintions Eq. (1) disappear and only
one of the four diagrams contributes. We will take Feymann gauge.
With the Feymann
rule in [5,6] one obtains the contribution from each diagram to the
fragmentaion function $D_{H/Q_1}(z)$. It is straightforward to calculate,
and in the calculation one meets the so called spin-sums, which are:
    $$ \eqalign {  u(p_1, s_1)\bar v(p_2,s_2) \chi_{s_1s_2}
       &= {\sqrt{m_1m_2}\over\sqrt{2} M } \gamma_5
      (\gamma\cdot P -M),\cr &\  {\rm for}\ H=\ ^1S_0\ {\rm state} \cr
    u(p_1, s_1)\bar v(p_2,s_2) (\epsilon_i(\lambda)\sigma_i\chi )_{s_1s_2}
       &=-{\sqrt{m_1m_2}\over\sqrt{2} M}
      (\gamma\cdot P+ M) \epsilon_p(\lambda)\cdot \gamma, \cr
        &\ {\rm for}\ H=\ ^3S_1\ {\rm state} \cr} \eqno (8) $$
Here $\epsilon_p(\lambda)$ is the 4-vector for the polarization of $H$
in the ${\bf P}\not= 0$ frame. Similar results for the spin sums in Eq.(8)
 and below can also be found in [7]. After a straightforward calculation we
obtain for the $H=\ ^1S_0$ bound-state:
  $$ \eqalign { D_{H/Q_1}(z) =& {2 \over 81\pi} {\vert R_0(0) \vert^2
      \over m_2^3}\alpha_s^2 {y_2z(1-z)^2 \over (1-y_1z)^6} \cr
     & \cdot \{ 6+18(1-2y_1)z+(68y_1^2-62y_1+15)z^2 \cr
     & \  +2y_1(-18y_1^2+17y_1-5)z^3+3y_1^2(1-2y_1+2y_1^2)z^4\}
      \cr } \eqno(9) $$
and for $H=\ ^3S_1$ bound-state:
  $$ \eqalign { D_{H/Q_1}(z,\lambda =0) =& {2 \over 81\pi} {\vert R_0(0)
\vert^2
      \over m_2^3}\alpha_s^2{y_2z(1-z)^2 \over (1-y_1z)^6} \cr
    & \cdot\{2-2(1+2y_1)z+(15-22y_1+16y_1^2)z^2 \cr
    & \ +2y_1(-5+7y_1-6y_1^2)z^3+3y_1^3(1-2y_1+2y_1^2)z^4\} \cr
    D_{H/Q_1}(z,\lambda =\pm 1 ) =& {2 \over 81\pi} {\vert R_0(0) \vert^2
      \over m_2^3}\alpha_s^2{y_2z(1-z)^2 \over (1-y_1z)^6} \cr
    & \cdot\{ 2-2(1+2y_1)z+(15-16y_1+10y_1^2)z^2 \cr
    & \ +2y_1(y_1-5)z^3+3y_1^2z^4\} \cr} \eqno(10)$$
Here $y_i=m_i/M$ for $i=1,2$ and $R_0(0)$ is the radial wavefunction at the
origin.
Comparing the fragmentation functions calculated in previous works[3,4]
we find agreement. Exchanging $y_1 \rightarrow y_2$ one can get
$D_{H/\bar Q_2}$. If we take $\bar Q_2=\bar Q_1$, then we obtain the
fragmentation functions for the quarkonia $(Q_1\bar Q_1)$.
\par
Now we turn to gluon fragmentaion function. If a hadron has $\bar Q_2
 =\bar Q_1$, then a gluon can decay into that hadron at the order of $g_s^4$.
In
this leading order there are also four Feymann diagrams depicted in Fig.2.
Since
Hadrons are colorless, there are always two gluons attached to the fermion
lines to bulid a color singlet. Because
charge-conjungation invariance
a gluon can not decay at this order into a hadron
with the quantum number $^3S_1$(i.e. $J^{PC}=1^{--})$.
Computing the contributions from the
diagrams we obtain the fragmentaion function for a $^1S_0$ hadron:
  $$ D_{H/G}(z)={\alpha_s^2 \over 24\pi}{\vert R_0(0) \vert^2
      \over m_1^3}\{(3-2z)z+2(1-z){\rm ln}(1-z) \} \eqno(11)$$
This is also in agreement with that obtained in [2]. With our results
calculated directly from Eq.(1) one can start with the defintions to
calculate the high order corrections. It should be pointed out that
the fragmentaion functions calculated here are at the scale $\mu =M$.
For the functions at an arbitrary scale $\mu$ one needs to solve
the correponding renormalization group equations and use the results
in Eq.(9,10,11) as initial conditions.
\par
With Eq.(9,10,11) we complete the calculations for the fragmentation
functions of a $S$-wave meson. The long-distance effect in these functions
are included in the nonrelativstic approach through the radial wave functions
at the origin, which can be calculated from potential models or can
be directly measured in leptonic decays. For a $P$-wave meson the
calculation is lengthy but still straightforward. Considering a meson to be
a $J^{PC}=1^{++}$ bound state of $(Q_1\bar Q_1)$, the creation operator
for this hadron is:
  $$ a^\dagger ({\bf P},\lambda ) = A_0 \sqrt{{3 \over 8\pi}}
     \int {d^3q\over (2\pi )^3} f_{1}(\vert {\bf q}\vert)
     { q_{i_1} \over \vert {\bf q}\vert}\epsilon_{i_2} (\lambda)
       (\sigma_{i_3} \chi)_{s_1s_2} \varepsilon_{i_1i_2i_3}
       a_{s_1}^\dagger (p_1)b_{s_2}^\dagger (p_2) \eqno(12)$$
and the spin sum used in the calculation:
      $$\eqalign { u(p_1,s_1)\bar v(p_2,s_2)&  q_{i_1}
 \epsilon_{i_2} (\lambda)
       (\sigma_{i_3} \chi)_{s_1s_2} \varepsilon_{i_1i_2i_3}= \cr
     &{1\over \sqrt{2}} {1\over 4M^3}
       (\gamma\cdot p_1+m_1)\varepsilon_{\mu\nu\sigma\rho}
       P^{\mu}\epsilon_p^{\nu}(\lambda)q_p^\sigma \gamma^\rho
        (\gamma\cdot P -M)(\gamma\cdot p_2-m_1)\cr } \eqno(13) $$
At order $\alpha_s^2$ a gluon can decay into the $J^{PC}=1^{++}$ meson, the
Feymann diagrams are same as in Fig.2. Keeping the leading order
at $\vert {\bf q}\vert$ we obtain:
  $$ \eqalign{ D_{H/G}(z,\lambda=0) & =
      {\alpha_s^2\over 108} {\vert R_1'(0)\vert ^2
   \over 4\pi m_1^5} \{ {z\over (1-z)}(1-z+2z^2)\} \cr
  D_{H/G}(z,\lambda=\pm 1) & =
      {\alpha_s^2\over 108} {\vert R_1'(0)\vert ^2
   \over 4\pi m_1^5} \{ {z\over (1-z)}(1-z+z^2)\} \cr }
    \eqno(14)$$
The $R_1'(0)$ is the derivative of the radial wavefunction at the origin.
The results in Eq.(14) are divergent when $z\rightarrow 1$. The reason
is that the gluon exchanged between the quarks in Fig.2 can move collinearly
along the quarks, if we take the nonzero leading order at
 $\vert {\bf q}\vert$. Such singularity is actually well known as the
singularity in the zero-binding limit[8], appearing in the calculations
of the hadronic decay widths of a $P$-wave meson. The situation is
similar, if we transpose the diagrams in Fig.2 and calculate the gluon
distribution of the hadron. The singularity at $z=1$ means that the
long-distance effect is present not only in the $R'_1(0)$ but also in the
remaining parts which we calculated perturbativly and hence it prevents
the naive use of perturbation thoery for $z$ near to 1. Recent progress[9]
on the $P$-wave hadron decay shows that the long-distance effect can be
clearly seperated from the short-distance effect by a {\sl new} factorization
theorem[9] for the decay, where one realizes that any meson is a
superpostion of many components:
  $$ \vert M >=\Psi_{Q\bar Q}\vert Q\bar Q >
                 +\Psi_{Q\bar Q G}\vert Q\bar QG> +\cdots
       \eqno(15)$$
and the configuration of $\vert Q\bar QG>$ plays an important roll in the
decay. In the hadronic decay of a $P$-wave meson one should also
take account on the $\vert Q\bar QG>$ configuration and the decay width
can be written in the {\sl new} factorization form, where instead of one,
two parameters represent the long-distance effect and the remaining parts can
be calculated perturbativly without the singularity. To apply this idea
in our case, one needs to know the wavefunction $\Psi_{Q\bar Q G}$ in Eq.(15),
which can not be obtained through the nonrelativstic approach. A detailed
study is needed. Here we recall a phenomenological treatment for the
singularity at $z=1$. Similar treatment of the decay was also used in [8,10].
In calculating the contribution from the diagrams to $D_{H/G}$
we integrated the transverse monmentum $\vert {\bf p_T}\vert$,
which is carried by the exchanged gluon, from zero to infinity. However,
the meson is not a point-like particle and it has a spatial extension.
The extension in the transverse direction is of the order $R \approx 1/M$.
If $\vert {\bf p_T}\vert$ is smaller than $1/R$, one should treat the
gluon as a part of the meson. From this point of view, the integral over
$\vert {\bf p_T}\vert$ should be
 from a nonzero $\vert {\bf p_T}\vert _{\rm min}\approx 1/R$
to $\infty$. We introduce
a parameter $\beta$:
     $$\vert {\bf p_T}\vert_{\rm min} =\beta M \eqno(16) $$
If the picture given above is correct, $\beta$ should be around 1. Here we
take it as a free parameter. With Eq. (16) we obtain:
  $$ \eqalign { D_{H/G}(z,\lambda=0) &= {\alpha_s^2\over 108}
            {\vert R_1'(0)\vert ^2
                   \over 4\pi m_1^5}((1-z)^2+z^2\beta^2)^{-3} \cr
     & \cdot \{ z(-2z^7+11z^6-26z^5+35z^4-30z^3+17z^2-6z+1) \cr
    & +3\beta^2z^4(-2z^4+7z^3-9z^2+5z-1)+3\beta^4z^5(-2z^3
       +4z^2-3z+1) \} \cr
       D_{H/G}(z,\lambda=\pm 1) &= {\alpha_s^2\over 108}
            {\vert R_1'(0)\vert ^2
                   \over 4\pi m_1^5}((1-z)^2+z^2\beta^2)^{-3} \cr
     & \cdot \{ z(-z^7+6z^6-16z^5+25z^4-25z^3+16z^2-6z+1) \cr
     & +3\beta^2 z^3(-z^5+4z^4-7z^3+7z^2-4z+1) \} \cr
       }\eqno(17) $$
The results in Eq. (17) are finite at $z=1$ and for $z\rightarrow 0$
the expressions approach those given in Eq. (14). In Eq.(17) there
are two parameters which need to be determined, the $R'_1(0)$ can be obtained
from potential models or measured elsewhere, while $\beta$ is unknown
in principle. However, in the fragmentation functions the dependence
on the momentum fraction $z$ is predicted by the perturbative
calculations. It should be stressed that it is important to know
the fragmentation functions for the $P$-wave meson or quarkonia. If we take
$Q_1$ as a c quark, the quarkonia  then corresponds to a $\chi_{c1}$
meson and $\chi_{c1}$ can substantially decay into $J/\psi$ via
radiative decay. Hence,  a theoretical prediction of  $\chi_{c1}$
production via fragmentation is very important for the prediction
for the $J/\psi$ production(for example see [11] and the references
cited there). \par
The defintions given in Eq.(1) are for unpolarized partons. One can
generalize the definitions to polarized partons and then obtain
fragmentation functions for a polarized parton. In this way
one can study through heavy meson production
the polarization properties of the quarks and gluons produced in high energy
collisions, where one expects for large $P_t$ processes that the
heavy mesons are produced dominantly via fragmentation[2].
\par
To summarize: We have shown in this letter that the fragmentation
functions for a $S$-wave meson composed of two heavy quarks
can be calculated directly from the definitions given by Collins and Soper
in [5] and the results obtained are same as those obtained by other
means. A new fragmentation function for a $P$-wave meson is also obtained,
but in contrast to a $S$-wave meson, there is an extra parameter which
needs
to be estimated. From the naive picture given above this parameter should be
around 1. Finally within the defintions one can conveniently study higher
order corrections to the fragmentaion functions.
\par\vskip 1cm
{\bf Acknowledgement:}\par
The author would like to thank Dr. M. Thomson for reading the manuscript
carefully. This work is supported in part by the Australian Research
Council.

\vfil\eject
\centerline{\bf References}\par
\noindent
[1] Chao-Hsi Chang and Yu-Qi Chen, Phys. Rev. D46 (1992) 3845
\par\noindent
\ \ \ \  C.R. Ji and F. Amiri, Phys. Rev. D35 (1987) 3318
\par\noindent
[2] E. Braaten and  T.C. YUan, Phys. Rev. Lett. 71 (1993) 1673
\par\noindent
[3] A.F. Falk, M. Luke, M.J. Savage  and M.B. Wise, Phys. Lett. B312 (1993)
486
\par\noindent
[4] E. Braaten, K. Cheung and  T.C. Yuan, Phys. Rev. D48 (1993) 4230
\par\noindent
[5] J.C. Collins and D.E. Soper, Nucl. Phys. B194 (1982) 445, Nucl. Phys.
B193 (1981) 381
\par\noindent
[6] J.C. Collins, D.E. Soper and G. Sterman, in Perturbative Quantum
Chromodynamics, edited by A.H. M\"ulller, World Scientific, Singapore,
1989.
\par\noindent
[7] B. Guberina, J.H. K\"un, R.D. Pecci and R. R\"uckl, Nucl. Phys.
B174 (1980) 317
\par\noindent
[8] R. Barbieri, R. Gatto and E. Remiddi, Phys. Lett. B61 (1976) 465
\par\noindent
\ \ \ \ R. Barbieri, M. Caffo and E. Remiddi, Nucl Phys. B162 (1980) 220
\par\noindent
[9] G.T. Bodwin, E. Braaten and G.P. Lepage, Phys. Rev. D46 (1992) R1914
\par\noindent
[10] W. Kwong, P.B. Mackenzie and R. Rosenfeld and J.L. Rosner,
  Phys. Rev. D37 (1988) 3210
\par\noindent
[11] M.L. Mangano, invited talk presented at the 9th Topical Workshop
on Proton--Antiproton Collider Physics, 18--22 October 1993, Tsukuba,
Japan, Preprint IFUP--TH 60/93
\par\noindent
\vfil\eject
\centerline{Figure Captions}\par\vskip 1cm
Fig.1. The Feymann diagrams for a quark to decay into a hadron. \par
Fig.2. The Feymann diagrams for a gluon to decay into a hadron.\par
\vfil\eject
\centerline{\bf Fig. 1}
\par\vfil\eject
\centerline{\bf Fig. 2}
\vfil
\end